\documentclass[11pt,dvips]{article}

\usepackage{epsfig,times} 
\usepackage{picinpar}
\usepackage{wrapfig}
\usepackage{floatflt}
\makeatletter
\renewcommand{\section}{\@startsection{section}{1}{0in}
	{0.4\baselineskip}{0.1\baselineskip}{\Large\bf}}
\renewcommand{\subsection}{\@startsection{subsection}{2}{0in}
	{0.25\baselineskip}{-\baselineskip}{\large\bf}}
\renewcommand{\subsubsection}{\@startsection{subsubsection}{3}{0in}
	{0.1\baselineskip}{-\baselineskip}{\normalsize\bf}}
\makeatother
\begin{document}

%
\makeatletter\newcommand{\ps@icrc}{
\renewcommand{\@oddhead}{\slshape{OG.2.1.01}\hfil}}
\makeatother\thispagestyle{icrc}
%
%

\begin{center}
%
{\LARGE \bf The energy spectrum of high energy gamma rays from Mrk501 by 
	the stereoscopic analysis}
\end{center}

\begin{center}
%
%
{\bf The Utah Seven Telescope Array collaboration}
\end{center}
\begin{center}
{\bf 	T.Yamamoto$^{1}$, 
	N.Chamoto$^{2}$, M.Chikawa$^{3}$, S.Hayashi$^{2}$, Y.Hayashi$^{4}$, N.Hayashida$^{1}$,
	K.Hibino$^{5}$, H.Hirasawa$^{1}$, K.Honda$^{6}$, N.Hotta$^{7}$,
	N.Inoue$^{8}$, F.Ishikawa$^{1}$, N.Ito$^{8}$, S.Kabe$^{9}$,
F.Kajino$^{2}$, T.Kashiwagi$^{5}$, S.Kakizawa$^{19}$, S.Kawakami$^{4}$, Y.Kawasaki$^{4}$,
     N.Kawasumi$^{6}$, H.Kitamura$^{16}$, K.Kuramochi$^{11}$, E.Kusano$^{9}$,
 E.C.Loh$^{12}$, K.Mase$^{1}$, T.Matsuyama$^{4}$, K.Mizutani$^{8}$, Y.Morizane$^{3}$,
    D.Nishikawa$^{1}$, M.Nagano$^{18}$, J.Nishimura$^{13}$, T.Nishiyama$^{2}$,
	M.Nishizawa$^{14}$, T.Ouchi$^{1}$, H.Ohoka$^{1}$, M.Ohnishi$^{1}$, S.Osone$^{1}$,
	To.Saito$^{15}$, N.Sakaki$^{1}$, M.Sakata$^{2}$, M.Sasano$^{1}$,
    H.Shimodaira$^{1}$, A.Shiomi$^{8}$, P.Sokolsky$^{12}$, T.Takahashi$^{4}$,
	S.F.Taylor$^{12}$, M.Takeda$^{1}$, M.Teshima$^{1}$, R.Torii$^{1}$, M.Tsukiji$^{2}$,
	Y.Uchihori$^{16}$,  Y.Yamamoto$^{2}$, K.Yasui$^{3}$, S.Yoshida$^{1}$,
	H.Yoshii$^{17}$, and T.Yuda$^{1}$
}\\
\vspace{1.0ex}
{\it $^{1}$Institute for Cosmic Ray Research, University of Tokyo, Tokyo 188-8502, Japan\\
\it $^{2}$Department of Physics, Konan University, Kobe 658-8501, Japan\\
\it $^{3}$Department of Physics, Kinki University, Osaka 577-8502, Japan\\
\it $^{4}$Department of Physics, Osaka City University, Osaka 558-8585, Japan\\
\it $^{5}$Faculty of Engineering, Kanagawa University, Yokohama 221-8686, Japan\\
\it $^{6}$Faculty of Education, Yamanashi University, Kofu 400-8510, Japan\\
\it $^{7}$Faculty of Education, Utsunomiya University, Utsunomiya 320-8538, Japan\\
\it $^{8}$Department of Physics, Saitama University, Urawa 338-8570, Japan\\
\it $^{9}$High Energy Accelerator Research Organization (KEK), Tsukuba 305-0801, Japan\\
\it $^{10}$Department of Physics, Kobe University, Kobe 657-8501, Japan\\
\it $^{11}$Faculty of Science and Technology, Meisei University, Tokyo 191-8506, Japan\\
\it $^{12}$Department of Physics, University of Utah, Utah 84112, USA.\\
\it $^{13}$Yamagata Academy of Technology, Yamagata 993-0021, Japan\\
\it $^{14}$National Center for Science Information System, Tokyo 112-8640, Japan\\
\it $^{15}$Tokyo Metropolitan College of Aeronautical Engineering, Tokyo 116-0003, Japan\\
\it $^{16}$National Institute of Radiological Sciences, Chiba 263-8555, Japan\\
\it $^{17}$Department of Physics, Ehime University, Matsuyama 790-8577, Japan\\
\it $^{18}$Department of Applied Physics and Chemistry, Fukui University of Technology, Fukui 910-8505, Japan\\
\it $^{19}$Department of Physics, Shinshu University, Matsumoto 390-8621, Japan\\
}
\end{center}

\begin{center}
{\large \bf Abstract\\}
\end{center}
We have developed the energy measurement method which based on 
stereoscopic observation with multiple telescopes for TeV gamma rays. 
Energy resolution obtained by this method was 23\%.
The energy spectrum of the gamma-ray flares 
of Markarian 501 in 1997 was also obtained using this techniques.
We have confirmed the bending or the cutoff of the energy spectrum around 
several TeV.

\vspace{-0.5ex}
%
%
%

\vspace{1ex}

%
%
\section{Introduction:}
\label{intro.sec}
Precise measurement of the gamma-ray energy spectrum in the TeV region from nearby
extra-galactic objects is one of the important objectives. We expect cutoff in their
energy spectrum by the interaction with the infrared photons in the inter galactic space.
The cutoff energy depends on the distance of the objects and density of infrared photons
(Stecker and M.Salamon 1997).
If we can measure the cutoff energy precisely as a function of redshift z, 
we can give constraints on the Hubble constant and the infrared photon density experimentally. 
The maximum energy of the electron acceleration in the AGN jet is another interesting topic.
It will represent the physics state of the jet and the super massive black hole which is 
supposed to exist at the center of the AGN.

The most promising way to measure the gamma-ray energy in the TeV region is the 
stereoscopic observation of the Cherenkov light from the air showers 
event by event. With this method, the axis of air shower, the arrival direction,
the intersection of the axis and the ground (core location) and the amount of Cherenkov
photon are detected more precisely. Making use of these advantages, we tried to
calculate the differential energy spectrum of TeV gamma rays.

Using the Monte Carlo Simulation, we derived the relation between the primary 
energy of gamma rays and several parameters, ADC value(ADCsum), the 
zenith angle, and the core distance which is the distance between the shower axis and the telescope.
In this procedure, we compared parameter distributions of experimental
data with those of simulation data and we found good agreement. 
Using this method, we have 
determined the differential energy spectrum of the gamma rays from Mrk501.

\section{Experiment:}
\label{obs.sec}
The Utah Seven Telescope Array has been in operation at Cedar
Mountain(1,600 m a.s.l.), Dugway in Utah 
($40.33^{\circ}$ N, $113.02^{\circ}$ W). 
Each telescope is arranged at the grid of a regular 
hexagon and at the center with a separation of 70m.
We have started the operation with three telescopes since March, 1997.

Each telescope has a 3 m diameter dish with
nineteen hexagonal segment mirrors and total effective mirror area of 
6 $m^2$.
The 256 channel camera with 0.25 degree pixels is mounted on the 
focal plane of each telescope. 
This camera is made of multi-anode photomultipliers(MAPMT) having 4 pixels. 

Details of this experiment are reported in OG.4.3.25.

\section{Analysis:}
\label{resul.sec}
In order to determine primary energies, directions and core positions
from images of Cherenkov light of air showers
generated by TeV gamma rays and cosmic rays, we used a simulation
based of the code of CORSIKA.



From shower images on the cameras of multiple telescopes, image parameters are
calculated and a shower axis is obtained.   
A detector-shower plane which includes the position of the telescope and the 
shower axis is determined for each telescope.
Then we can determine the intersection line of these planes, which corresponds 
to the three-dimensional shower axis.
Core location of the shower is determined as the intersection point of the shower
axis and the ground plane. Figure 1 shows these procedures and the resolution.
Position resolution obtained by this method is 12 m in which 50 \% of the total 
events are included.

The relation between the total ADC counts and the primary energy of gamma rays is
estimated as a function of zenith angle and core location.
Based on this relation, the primary energy of gamma rays is determined
event by event. 
The energy resolution obtained from this analysis is estimated to be 23 \% (Figure 2.a).

\section{Energy spectrum:}
\label{spect.sec}
For the analysis in this paper, we use all of the events that were detected by two 
telescopes. In the case of the observation using three telescopes, there are three 
combinations of two telescopes. It means the efficiency of the observation using 
three telescopes is three times higher than two telescopes.
The data used in this paper was acquired from the beginning of April to the end of 
the July 1997. In April, two telescopes were operated and one more telescope joined to the
observation since May. Total observation time was 137.6 hours.

Figure 3.a shows the primary energy distribution of the detected air showers 
assuming the primary particles are gamma rays.
We calculate the energy distributions in the on-source and 
in the off-source region, respectively. With this figure, the energy 
distribution of gamma rays from Mrk501 can be estimated. 
Subtracting off-source events from on-source events,
the energy distribution of the gamma rays is obtained.
Taking into account the effective area and the observation time, 
differential energy spectrum of the gamma rays 
from Mrk501 is calculated. Figure 3.b)

\section{Conclusion:}
We have developed energy determination method for gamma rays
using stereoscopic technique.
Accuracies of core location determination and energy determination are 12m and 23\%,
respectively.
Note that the image parameters like WIDTH and LENGTH are
not used except to determine the image axis in this analysis.
We have obtained the differential energy spectrum 
of the gamma-ray flare of Mrk501 in 1997.
The flux of the gamma rays is well represented by a differential power spectrum 
with an index of -2.5 between 700 GeV and 3 TeV 
and the steepening effect seems to appear above several TeV. 


\hspace{0.5cm}{\bf Acknowledgments}

This work is supported in part by the Grants-in-Aid for Scientific Research 
(Grants \#0724102 and \#08041096) from the Ministry of Education, Science
and Culture. The authors would like to thank the people at Dugway for the 
help of observations.

\vspace{1ex}
\begin{center}
{\Large\bf References}
\end{center}
%
Hayashida, H. et al 1998, ApJ 504, L71.\\
Nishikawa, D. et al 1999,Proc.26$^{th}$ ICRC(Salt Lake city,1999)OG.2.1.17\\
Yamamoto, T. et al 1999,Proc.26$^{th}$ ICRC(Salt Lake city,1999)OG.2.1.25

\begin{figure}[bht]
\begin{center}
\epsfig{file=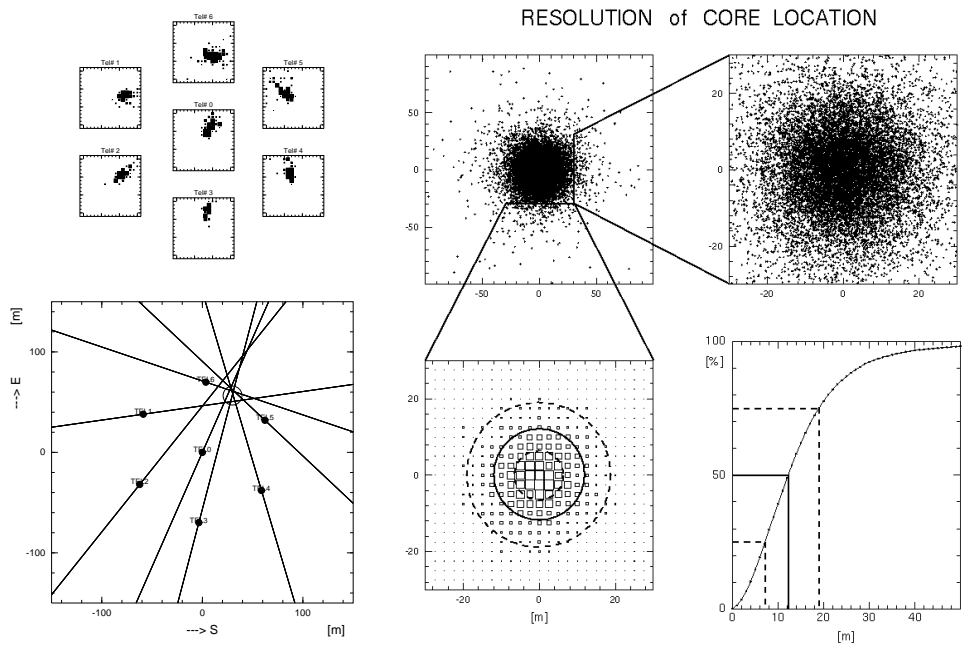,width=5.3in}
\caption{(a) The Left top figure shows the images of air shower initiated by 
	     gamma ray made by Monte Carlo simulation.
	     Left bottom panel shows the arrangement of the telescopes, core location
	     and 
	     line of intersection of the shower plane
	     and ground. The real core location is given by the white circle.
        (b)  Right panel shows the resolution of core location calculated by
	     the Monte Carlo simulation. The center of each panel
	     is the real core location. 
	     The estimated core location from the shower
	     plane are plotted in the area of 200 m and 60 m.
	     The bottom right panel shows the number of the estimated core locations 
	     in each area. It shows boundary in which 50\% of the total events are contained
		is 12 m.
}
\label{fig:s1}
\end{center}
\end{figure}

\begin{figure}[bht]
\begin{center}
\epsfig{file=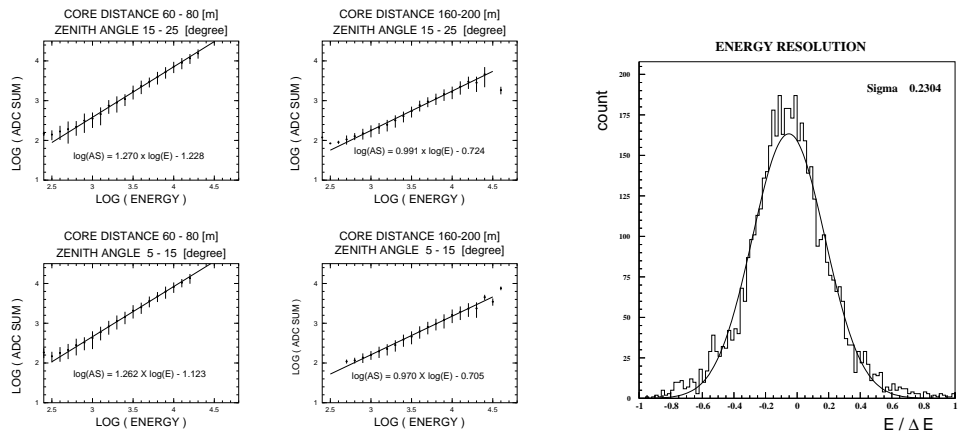,width=5.2in}
\caption{(a) Left panel shows the relation between primary gamma-ray energy
             and ADC sum in each core distance and each zenith angle. 
         (b) Right panel shows the energy resolution of this analysis
             calculated by the Monte Carlo simulation.
	     According to this figure, the energy resolution of this 
             analysis is shown to be 23 \%.
}
\label{fig:s2}
\end{center}
\end{figure}

\begin{figure}[bht]
\begin{center}
\epsfig{file=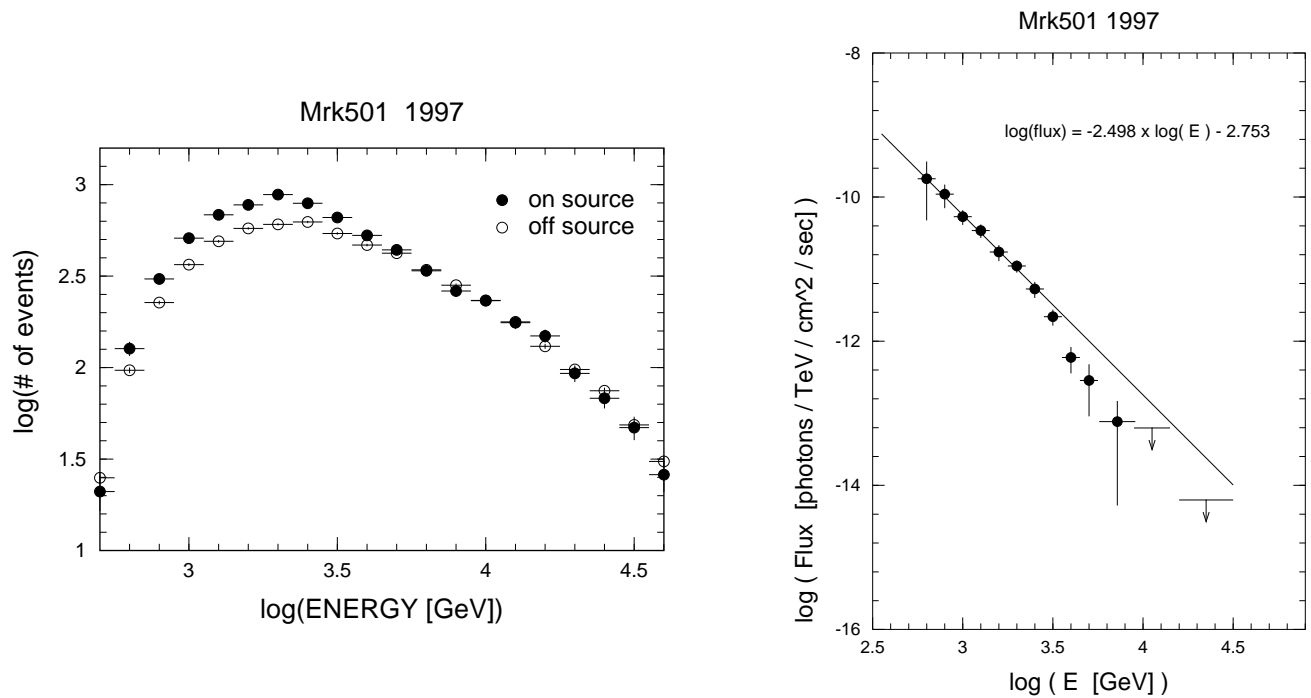,width=7.2in}
\caption{(a) Left panel shows the distribution of the primary energy of the air 
             showers which are detected by more than two telescopes.
             The on source area and the off source area are presented.
         (b) Right panel shows the result of the calculation of the differential
             energy spectrum of the gamma rays from Mrk501 observed in 1997 
	     obtained by the stereo analysis.
}
\label{fig:s3}
\end{center}
\end{figure}

\end{document}